\documentclass[a4paper,10pt]{article}
\usepackage{graphicx}
\usepackage{dcolumn}
\usepackage{bm}
\usepackage{amsmath}
\usepackage{amssymb}

\title{Spin Glass and ferromagnetism in disordered Cerium compounds}

\author{S. G. Magalhaes\footnote{ggarcia@ccne.ufsm.br} $^{,a}$, F. M. Zimmer$^a$, P. R. Krebs$^b$, B. Coqblin$^c$%
\\
\\
{$^a$\normalsize{\it Lab. de Mec\^anica Estat\'\i stica e Teoria da Mat\'eria Condensada,}}
\\
{\normalsize\it Dep. F\'\i sica, UFSM, 97105-900 Santa Maria, RS, Brazil}
\\
{$^b$\normalsize\it Instituto de F\'\i sica e Matem\'atica,      
Universidade Federal de Pelotas,}         
\\
{\normalsize\it Caixa Postal 354, 96010-900 Pelotas, RS, Brazil}
\\
{$^c$\normalsize\it Laboratoire de Physique des Solides,
Universit\'e Paris-Sud,}
\\
{\normalsize\it b\^atiment 510, 91405 Orsay, France}}

\date{}
\begin{document}
\maketitle

\begin{abstract}
The competition between spin glass, ferromagnetism and Kondo effect { is} 
analysed 
{ here} 
in a Kondo lattice model 
{ with an}
inter-site { random} coupling $J_{ij}$ between the localized magnetic 
moments 
given by a 
{ generalization of the}
Mattis model \cite{Mattis} 
which represents 
an interpolation between ferromagnetism and a highly disordered spin glass. 
Functional integral techniques with Grassmann fields have been used to obtain the partition function.
The static approximation and the replica symmetric ansatz have also been used. 
The solution of the problem is presented as a phase diagram 
{ giving} 
$T/{J}$ {\it versus} $J_K/J$ where $T$ is the temperature, $J_{K}$  and ${J}$
are the strengths of the intrasite Kondo and the intersite random couplings, respectively. 
If $J_K/{J}$ is small, 
when temperature is decreased, there is a second order transition from a paramagnetic to 
a spin glass phase.
For lower $T/{J}$, a first order transition 
appears between the spin glass phase and a region where there are Mattis states which are
thermodynamically equivalent to the ferromagnetism.  
For very low ${T/{J}}$, the Mattis states become 
stable.  On the other hand, 
{ it is found as solution a}
Kondo state for { large} $J_{K}/{J}$ values. These 
results can improve the theoretical description of the well 
known experimental phase diagram of $Ce$$Ni_{1-x}$$Cu_{x}$ \cite{Gomez-Sal1,Gomez-Sal2, Gomez-Sal3, Espeso}. 
\end{abstract}

\section{Introduction}

The properties of many cerium or uranium compounds are well described by the Kondo-lattice model, 
with strong competition between the Kondo effect on each site and the Ruderman-Kittel-Yosida-Kasuya (RKKY) interaction 
between magnetic atoms at different sites. The role of disorder has been studied in disordered 
alloys containing cerium or uranium and
different theories have been proposed. In the Kondo disordered model (KDM) \cite{Miranda1, Miranda2}, 
disorder produces a broad distribution of Kondo temperatures and can be responsible for the deviation 
from the Fermi Liquid behavior found in some heavy fermion systems. Another theoretical 
approach is the Magnetic Griffths phase \cite{Castro-Neto}, where  
fluctuations of the magnetic clusters can produce Griffiths-McCoy singularities close to a 
Quantum Critical Point (QCP). On the other hand, we have studied within a mean field approximation 
the phase diagrams observed in disordered heavy fermion systems showing Kondo, spin glass and
magnetically ordered phases \cite{Alba1,Magal1,Magal2} and we will discuss these models later on. 
Earlier studies have also suggested that a spin glass transition near a QCP  could lead to a Non Fermi Liquid 
(NFL) behavior \cite{Georges}. 
{ Our paper is an attempt} to improve  the theoretical 
description of the spin glass-Kondo-ferromagnetic competition 
in order to obtain a better agreement with the experimental situation of disordered
cerium or uranium heavy fermion systems. 

Spin glass and Kondo state have been observed together in several Cerium alloys like 
$CeNi_{1-x}Cu_{x}$ \cite{Gomez-Sal1, Gomez-Sal2}, $Ce_{2}Au_{1-x}Co_{x} Si_{3}$ \cite{Majundar} and 
in some disordered Uranium alloys such as 
$UCu_{5-x}Pd_{x}$ \cite{Volmer} or $U_{1-x}La_{x}Pd_{2}Al_{3}$ \cite{Zapf}. The first  
case has been studied by bulk methods (see { Refs}. \cite{Gomez-Sal1, Gomez-Sal2}) and more 
recently by $\mu$SR 
spectroscopy \cite{Gomez-Sal3} which gives local information about the spins configurations. The bulk probes 
have shown a presence of antiferromagnetic 
phase for low $Ni$ content (for instance $x = 0.9$). In the region 
$x \lesssim 0.2$, the Kondo effect { becomes important}
producing magnetic moment reduction. For $0.8\gtrsim x \gtrsim 0.4$,  the same bulk 
probes have shown a presence of 
{ the} spin glass like state intermediate in temperature between a ferromagnetic order 
(at lower temperature) and paramagnetism (at higher temperature).  
However, the $\mu$SR 
spectroscopy has shown in the 
region $0.8 \gtrsim x \gtrsim 0.4$  a scenario favoring the presence of a 
inhomogeneous ``cluster spin glass" { (or called equivalently "cluster glass" in Ref. \cite{Gomez-Sal3})} rather than a standard spin glass.  
Quite recent measurements 
on the specific heat \cite{Espeso} 
{ has confirmed} the emergence of a spin glass-like state 
and a percolative evolution to 
a ferromagnetic order at low temperatures.

There has been a theoretical attempt \cite{Magal1} to build up a 
global phase diagram based on a Kondo lattice model with 
a random Gaussian inter-site 
coupling among the localized spins with mean $2 J_{0}/N$ and 
standard deviation $\sqrt{8 \tilde{J}^{2}/N}$  ($N$ is the number of sites). 
The spins operators have been given as bilinear combination of creation and 
destruction fermionic 
operators. The partition function has been found using path integral formalism 
within the static 
approximation and replica symmetry ansatz \cite{Alba1}.   
The results { have shown} 
that  ferromagnetism, spin glass 
and a mixed phase (a solution with non-zero magnetization below the Almeida-Thouless line) 
have been obtained for small $J_K/\tilde{J}$ values
and $J_{0}/\tilde{J}>1.46$  while a Kondo phase is obtained for 
large $J_{K}/\tilde{J}$ values.  

However, the calculated spin glass freezing temperature ($T_f$) is lower 
than the Curie temperature ($T_c$) in this highly frustrated model. 
Even the { transition temperature to the} 
mixed phase { is} 
always below the onset of the ferromagnetic order {at $T_{c}$}. 
Thus, our previous model \cite{Magal1} gives a ferromagnetic 
{ transition temperature ($T_{c}$)} 
above the spin glass { transition temperature ($T_{f}$)}, 
in contrast with the experimental situation of $CeNi_{1-x}Cu_{x}$ \cite{Gomez-Sal1,Gomez-Sal2,Gomez-Sal3,Espeso} 
alloys where the ferromagnetic phase 
is always the lowest one. That would be a clear indication that 
Gaussian distributed random couplings as in
the Sherrington-Kirkpatrick (SK) model  \cite{SK} is not adequate to describe the frustration present in that alloy.

One important point in the set of experimental works has been to clarify the
effect of the disorder in the $CeNi_{1-x}Cu_{x}$ \cite{Gomez-Sal1,Gomez-Sal2,Gomez-Sal3,Espeso}. When
$Cu$ is randomly replaced by $Ni$ in that alloy, not
only the cell volume is modified, but also the
number of conduction electrons which makes the competition between the
RKKY and Kondo effect complex for  that particular alloy \cite{Espeso}.
As a consequence, one localized spin at any site can be 
subject to a set of effective local magnetic fields, 
which results in the complicated combination of states previously cited.

In the Mattis model \cite{Mattis}, which has been proposed as a solvable model to the 
spin glass problem,  the bonds joining the localized spins 
have been  defined as separable random variables $\xi_i$. This model could
allow to gain some insight about the local effects of disorder as long 
as it would be possible to construct local applied fields dependent on the random variable $\xi_{i}$. Unfortunately, 
at zero { magnetic } field, a gauge transformation of the Ising spins classical variables leads  Mattis 
model free energy to behave rather as the usual ferromagnet \cite{Binder,Fradkin}.  Therefore, this model 
is trivially disordered in the sense that { it     } is unable to produce the essential component of the spin glass 
which is frustration.
Nevertheless, the generalization of 
the Mattis model \cite{Provost,VanHemmen} has proved 
to be an interesting alternative. In this model the coupling between spins are given by 
\begin{equation}
J_{ij}=\frac{1}{N}\sum_{\mu\nu}J_{\mu\nu}\xi_i^{\mu}\xi_j^{\nu} 
\label{Jij},
\end{equation}
where $\xi_i^{\mu}=\pm 1$ ($\mu=1,2,...,p$; $i=1,2,...,N$) are independent random distributed variables. 
For the classical Ising model, if $\mu=\nu=1$, the original 
Mattis model \cite{Mattis} is recovered. However, if $J_{\mu\nu}=J\delta_{\mu\nu}$ and p=N  with 
the $N^{2}$ random variables $\xi_i^{\mu}$ having mean zero and variance one, in the 
limit of N large, $J_{ij}$ tends to a Gaussian variable with mean zero and variance  
$N^{-1/2}J$ as in the SK model \cite{SK}. Therefore, { we} can consider this  
model as an interpolation between ferromagnetism and highly disordered spin glass \cite{Amit1}. 

{ An important particularity of this}  model (see Eq. (\ref{Jij})) is
$J_{\mu\nu}=J\delta_{\mu\nu }$ which has been 
used in a different context{, i. e.} the statistical 
mechanics theory of { complex systems} \cite{Amit1} 
using classical Ising spins.
In this problem, randomness effects can be better understood at $T=0$ 
temperature when 
the local field applied 
in a particular spin  given by $h_{i}=\sum_{i\neq j}J_{ij}S_{j}$ {  is analysed}  
\cite{Amitnew}. In the state  $S_{i}=\xi_{i}^{1}$  (choosing ${J}=1$), the local 
field becomes $h_{i}=\xi_{i}^{1}(1+\delta_{i})$, where $\delta_{i}$ is 
a random variable with variance $<\delta_{i}^{2}>_{\xi}=\frac{p-1}{N}$. Two 
situations can be identified when $N\rightarrow \infty$ ($N$ is the number 
of sites). If $p$ is finite, the spin is perfectly aligned with $\xi_{i}^{1}$. 
However, if $p$ increases linearly with $N$ ($p=a N$), the term $\delta_{i}$ 
can become important and the alignment can be destroyed. This random 
component of the local field can be a source of frustration and, in that sense, the ratio $\sqrt{N/p}$ is the analog 
{ of} 
$J_{0}/\tilde{J}$ \cite{Amit1}. When temperature 
is turned over, there is an additional mechanism to avoid the aligning. 

Actually, the thermodynamics of { the generalized classical Mattis model}  can be described by a { mean field theory \cite{Amit1,Amit2,Amit3,Amitnew}} in terms of the parameter
$a=p/N$. For a particular value called $a_{c}$, 
two clearly distinct regimes can be identified. 
When $a>a_{c}$, the frustration is dominant { below a certain temperature $T_{f}$} 
leading 
the problem to a spin glass behavior. 
Nevertheless, when $0<a<a_{c}$, a much more 
complex scenario can appear depending on the temperature. For instance, 
{ it can be found as solution not only a spin glass phase, but also} 
Mattis states (which corresponds to the stable aligning between 
$\xi$'s and spins)
{ with} a first order transition between them. 
These Mattis states have the same thermodynamics as the ferromagnetic phase  
\cite{Amit1,Amit2,Amit3,Amitnew}.

The mean field 
description of the previous situation  
introduces 
the order parameter
$m^{\mu}=\frac{1}{N}\sum_{i}\xi_i^{\mu}<S_{i}>$ which gives a measure of the difference 
between the configurations of the set  $\{\xi_i^{\mu}\}$
and the spins. In that approach, 
{ one particular solution can be, 
for example, only $m^{1}$ with a possible non zero value, 
while the remaining 
$m^{\mu}$ 
($\mu \neq 1$) 
are {of order $1/\sqrt{N}$}. 
This choice corresponds to the situation 
where the spins 
can be 
perfectly aligned only with $\xi_{i}^{1}$.} 
{ Nevertheless}, 
these $m^{\mu}$'s {($\mu \neq 1$)} still have a role in the problem 
yelding
a possible spin glass solution in the problem depending on the temperature and, particularly,
on the parameter $a$. 
Therefore, this approach would be mathematically 
convenient to apply in a problem 
where we would be interested in getting control about the degree of  frustration.

In this work, we consider the Kondo lattice model 
with 
a random intersite interaction between the localized spins where the coupling $J_{ij}$ 
is given by Eq. (\ref{Jij}).
That would allow us to investigate the competition between Kondo effect and 
magnetism combining   the approach of { Ref.} \cite{Alba1} with { Ref.} \cite{Amit2}. 
Therefore, the degree of frustration $a=p/N$ is a new parameter which together with $J_{K}/J$ 
($J_K$ is the strength of the intra-site Kondo coupling)
constitute the parameter space where the solutions can be located. 
In the { Ref.} \cite{Alba1}, the solutions for the order parameters  have been found only in the limit of strong 
frustration (the random Gaussian coupling) which corresponds to the $a>>a_{c}$ situation. We will show a more complex 
situation in the limit of weak frustration. For { small} $J_{K}/J$, there is an intermediate spin glass phase between 
paramagnetism and
{ the region where there are Mattis states which corresponds to the ferromagnetism}. 
Furthermore, the transition from the spin glass 
to the ferromagnetism is first 
order. Therefore, there is a 
large region in temperature where the ferromagnetic solution is thermodynamically metastable. { It becomes} stable 
at very low temperature. 
For large
$J_{K}/J$, we get  a Kondo state, as already discussed  in { Ref.} \cite{Alba1}.

It is not obvious that the properties of the classical Mattis model 
and its generalization can be extended to the quantum version 
of these models. However, it has been shown that the long range quantum Mattis model 
has the same qualitative  behavior of his classical counterpart 
\cite{Coolen}. Moreover, in the present fermionic spin glass mean field approach 
with the static approximation, which is 
{ reliable} at high temperature, the spin 
variables have the essential features of classical ones \cite{Alba2,Alba3}. 

It should be remarked that the extension of the method 
given in { Ref.} \cite{Amit2} to the present fermionic problem is not straightforward. 
In fact, the model developed here is different from that one 
introduced in { Refs.} \cite{Magal1, Alba1, Magal2}. Indeed, both 
models study the competition between Kondo 
effect and spin glass, with eventually an additional magnetic phase. 
But, the description of the spin glass onset 
is different in the two cases. In the present one, the parameter $a$ can tune 
the spin glass or the ferromagnetic component of the long range internal field {which 
has a completely different dependence on the spin glass order parameter as compared 
with Refs. \cite{Magal1, Alba1, Magal2}}. 
Actually, for large values of the parameter $a$, the Gaussian distribution of spin couplings 
is recovered. That corresponds to the strong frustration situation studied in { Ref.} \cite{Alba1}.   
Thus, the theoretical description is not 
simple, but the present model is able to 
give a more 
local description of the problem. Finally, as it will be discussed 
later on, we will obtain a ferromagnetic phase below 
the spin glass solution { in terms of temperature}, in contrast to the results  
of the previous model and in improved agreement with the experimental 
phase diagram of $Ce$$Ni_{1-x}$$Cu_{x}$ alloys.

On the other hand, we can say that the experimental situation of 
$Ce$$Ni_{1-x}$$Cu_{x}$ or similar alloys are also very complex. In fact, 
the so-called spin glass phase is more exactly a ``cluster { spin}  glass" 
which tends to a real ferromagnetic order by a percolative process 
when temperature decreases.

The outline of the paper is the following. In section II, 
the model is introduced and developed 
in order to get the free energy with the 
relevant order parameters for the problem. 
The results obtained are presented and 
discussed in section III, and we 
{ finished it up} with a conclusion in the last section.

\section{General Formulation}

The model is the Kondo lattice used previously to study the competition between
spin glass and Kondo effect \cite{Alba1}. The Hamiltonian is given by
\begin{equation}
H=\sum_{k,\sigma}\epsilon_{k} n^{c}_{k\sigma}+ \sum_{i,\sigma}\epsilon_{0} n^{f}_{i\sigma}
+H_{SG}
+J_{K}\sum_{i}
[\hat{S}^{+}_{f,i}\hat{s}^{-}_{i}
+\hat{S}^{-}_{f,i}\hat{s}^{+}_{i}]
\label{e2}
\end{equation}
where the sum is over the $N$ sites of a lattice. 

The term $H_{SG}$ corresponds to the intersite interaction between localized
spins, thus
\begin{equation}
H_{SG}=\sum_{i,j}J_{ij}\hat{S}^{z}_{fi}\hat{S}^{z}_{fj}
\label{e3}.
\end{equation}

The random coupling $J_{ij}$ present in the previous equation is given 
{ by}
Eq.
(\ref{Jij}) with $J_{\mu\nu}=\frac{J}{2}\delta_{\mu\nu}$ \cite{Amit1},
where $\xi^{\mu}_{i}=\pm 1$ are random independent variables which follow 
the distribution:
\begin{equation}
P(\xi^{\mu}_{j})=1/2\delta_{\xi^{\mu}_{j},+1}+1/2\delta_{\xi^{\mu}_{j},-1}
\label{e5}.
\end{equation}

The spin operators in Eq. (\ref{e2}) are defined (see { Refs.}
\cite{Magal1, Alba1, Magal2}) as bilinear combinations of the creation and
destruction operators for localized (conduction) fermions
$f_{i\uparrow}^{\dagger}$, $f_{i\downarrow}$ $(d_{i\uparrow}^{\dagger}$, 
$d_{i\downarrow}$) with the spin projection $\sigma=\uparrow$ or $\downarrow$:
%
$\hat{S}_{fi}^{+}=f_{i\uparrow}^{\dagger}f_{i\downarrow}$; $
\hat{S}_{fi}^{-}=f_{i\downarrow}^{\dagger}f_{i\uparrow}$; $
\hat{s}_{ci}^{+}=d_{i\uparrow}^{\dagger}d_{i\downarrow}$; $
\hat{s}_{ci}^{-}=d_{i\downarrow}^{\dagger}d_{i\uparrow}$;
%
\begin{equation}
{\hat{S}_{fi}^{z}}=\frac{1}{2}[f_{i\uparrow}^{\dagger}f_{i\uparrow}-
f_{i\downarrow}^{\dagger}f_{i\downarrow}]
\label{e6}.
\end{equation}
%

The chemical potential for the localized and conduction bands are $\mu_{f}$ and
$\mu_{c}$, respectively. As it has been done in { Ref.}
\cite{Magal1,Alba1,Magal2}, the energy $\epsilon_{o}$ is referred to $\mu_{f}$ and
$\epsilon_{k}$ is referred to $\mu_{c}$.

In the functional integral formalism, the partition function is expressed
using anticommuting Grassmann variables $\varphi_{i\sigma}(\tau)$ (related to
conduction electrons) and $\psi_{i\sigma}(\tau)$ (related to the localized
electrons) \cite{Alba1} as
\begin{align}
Z  = \int D(\psi^{\ast}\psi) D(\varphi^{\ast}\varphi)
\exp\left[A_{SG} + A_K \right.
\left.+ A_0(\psi^{\ast}\!,\psi)+A_0(\varphi^{\ast}\!,\varphi) \right] 
\label{e8}
\end{align}
where { in the static approximation \cite{Alba1} }
\begin{eqnarray}
A_0(\psi^{\ast}\!,\psi)  =\sum_{ij\sigma} \sum_{\omega}
\psi_{i\sigma}^{\ast}(\omega) \left[i\omega -\beta \varepsilon_{0}\right]
\delta_{ij}\psi_{j\sigma}(\omega),~ 
\\
A_0(\varphi^{\ast}\!,\varphi)  =  \sum_{ij,\sigma,\omega}
\varphi_{i\sigma}^{\ast}(\omega) \left[(i\omega -
\beta\varepsilon_{k})\delta_{ij} - t_{ij} \right] 
\varphi_{j\sigma}(\omega), 
\\
A^{stat}_K \approx \frac{J_K}{N} \sum_{i\sigma} \sum_{\omega}
\left[ \varphi_{i-\sigma}^{\ast}(\omega)
\psi_{i-\sigma}(\omega)\right]
\sum_{j\sigma}\sum_{\omega^{'}}\left[ \psi_{j\sigma}^{\ast}(\omega^{'})
\varphi_{j\sigma}(\omega^{'})\right],
\\
A^{stat}_{SG} ={\displaystyle \sum_{ij}} J_{ij} S_{i}S_{j}.~~~~~~~~~~~~~~~~~~~~~~~~~~~~
\label{e9}
\end{eqnarray}

with  
\begin{eqnarray}
S_{i}=\frac{1}{2}{\displaystyle \sum_{\sigma=\pm}}{\displaystyle \sum_{\omega}}\sigma \psi_{i \sigma}^{\dagger}(\omega)\psi_{i \sigma}(\omega).
\label{equ10}
\end{eqnarray}
In Eq. (\ref{equ10}), the Matsubara's frequencies { are given, as usual, by} $\omega=(2m+1)\pi$ with $m=0,\pm
1,\pm 2,\dots$. 

The problem is treated closely to the mean field approximation 
{ of}
{ Ref.}
\cite{Alba1}.
{ Therefore,} the
Kondo order parameter $\lambda_{\sigma}$ and its conjugate are introduced using
the integral representation of the delta function as
\begin{eqnarray}
\delta\left( N\lambda_{\sigma}-\sum_{\omega}\sum_{i=1}^{N}
\varphi_{i\sigma}^{*}(\omega)\psi_{i\sigma}(\omega) \right)
=\int\prod_{\sigma}\frac{dv_{\sigma}}{2\pi}\times\nonumber\\
\exp\left\{i\sum_{\sigma}v_{\sigma}\left[
N\lambda_{\sigma}^{*}-\sum_{\omega}\sum_{i=1}^{N}
\varphi_{i\sigma}^{*}(\omega)\psi_{i\sigma}(\omega)\right]\right\}
\label{e10}
\end{eqnarray}
{ where the presence of the order parameter $\lambda_{\sigma}$ is related 
to the formation of $d-f$ singlet throughout the whole lattice. This mean field 
order parameter is presently recognized to provide good description of the Kondo effect 
on each site \cite{Lacroix}}

Therefore, the resulting partition function becomes:
\begin{equation}
Z=\int\prod_{\sigma} d\lambda_{\sigma}^{\dagger} d\lambda_{\sigma}
\exp\left[-N\beta
J_{K}\sum_{\sigma}\lambda_{\sigma}^{\dagger}\lambda_{\sigma}\right]Z_{stat}
\label{e11}
\end{equation}
\noindent
where 
%
\begin{eqnarray}
Z_{stat} = \int D(\psi^{\ast}\psi) D(\varphi^{\ast}\varphi)
\exp\{A_0(\psi^{\ast},\psi)+A_{SG}^{stat}+A_0(\varphi^{\ast}\!,\varphi)
\nonumber\\
+  
\beta J_{K}\sum_{\sigma}
~[~\lambda_{-\sigma}^{\dagger}\sum_{j,\omega} 
\varphi_{j\sigma}^{\dagger}(\omega)\psi_{j\sigma}(\omega) 
+\lambda_{\sigma}\sum_{j,\omega}\psi_{j\sigma}^{\dagger}(\omega)
\varphi_{j\sigma}(\omega)~]~\}
\label{e12}~.
\end{eqnarray}
{ In fact, in this work the Kondo order parameter is considered $\lambda_{\sigma}=\lambda$ \cite{Alba1,
Magal1, Magal2}.} 

The integration over the Grassmann fields $\varphi^{\ast}$ and $\varphi$ { in Eq. (\ref{e12})}
can be performed, which results { in:}
\begin{equation}
\frac{Z_{stat}}{Z_{0}} = \int D(\psi^{\ast}\psi)
\exp\left[A_0^{eff}+A_{SG}^{stat}\right]
\label{e13}~
\end{equation}
\noindent
where  
\begin{equation}
A_{0}^{eff}\!=\!\sum_{ij\sigma}\sum_{\omega}\psi_{i\sigma}^{\ast}(\omega)
g_{ij}^{-1}(\omega)\psi_{j\sigma}(\omega),~
\label{e14}
\end{equation}
with
\begin{equation}
g_{ij}^{-1}(\omega)\!=\!(i\omega-\beta\epsilon_0)\delta_{ij}
-\beta^{2}J_{K}^{2}\lambda^{\ast}\lambda\gamma_{ij}(\omega)~
\label{e151}.
\end{equation}
The Fourier transform of Green's function $\gamma_{ij}(\omega)$ in Eq.
(\ref{e151}) is 
\begin{equation}
\gamma_{k}(\omega)=\frac{1}{i \omega -\beta\varepsilon_{0}-\beta\varepsilon_{k}}
\label{e16}
\end{equation}
In order to introduce the proper { set of} order parameters in { our} problem \cite{Amit2},
the action $A_{SG}^{stat}$ in Eq. (\ref{e9}) is written { using Eq. (\ref{Jij})}
to give
\begin{equation}
A_{SG}^{stat}=\frac{\beta J}{2N}\sum_{\mu=1}^{p}(\sum_{i}\xi_{i}^{\mu}S_{i})^2 -
 \frac{\beta J p}{2N} \sum_{i}(S_{i})^2
\label{eqq15}.
\end{equation}
where $S_{i}$ has been defined in Eq. (\ref{equ10}).

The free energy can be obtained following the replica method,
\begin{equation}
\beta f=2\beta J_{k}\lambda^{\ast}\lambda - \lim_{n\rightarrow 0}\frac{1}{Nn}
\left( \langle\langle Z_{stat}^{n}\rangle\rangle_{\xi}-1 \right)
\label{e17}~,
\end{equation}
\noindent
where $\langle\langle \cdots\rangle\rangle_{\xi}$ is the averaged over $\xi$'s.

The fundamental issue consists in the evaluation of the quadratic form present
in the first term of $A_{SG}^{stat}$. 
First, the sum over $\mu$ is
separated in two parts \cite{Amit2}:
%
$\sum_{\mu=1}^{p}=\sum_{\mu=s}^{p}+\sum_{\nu=1}^{s-1}~.$

It is possible to linearize the problem introducing $n\times p$ auxiliary fields
$m_{\alpha}^{\mu}$ and $m_{\alpha}^{\nu}$ ($\alpha$ is a replica index) which correspond to the 
parameter discussed in the previous section. Therefore, 
\begin{eqnarray}
\exp(A_{SG}^{stat}) = \exp\left[ -\frac{\beta J
p}{2N}{\displaystyle \sum_{\alpha=1}^{n}}{\displaystyle \sum_{i=1}^{N}}(S_{i}^{\alpha})^2\right] 
\times
\nonumber\\
\int_{-\infty}^{\infty} Dm_{\alpha}^{\mu}                 
\exp\left\{
\beta J N\sum_{\nu=1}^{s-1}\sum_{\alpha}\left[-\frac{1}{2}(m_{\alpha}^{\nu})^2
+\right.\right. 
\left.\left.
\frac{1}{N}\sum_{i} \xi_{i}^{\nu} S_{i}^{\alpha} m_{\alpha}^{\nu} \right]\right\} 
\int_{-\infty}^{\infty}  Dm_{\alpha}^{\nu}                           
\times\nonumber\\ \exp\left\{
\beta J N \sum_{\mu=s}^{p}\sum_{\alpha}\left[-\frac{1}{2}(m_{\alpha}^{\mu})^2 +
\frac{1}{N}\sum_{i} \xi_{i}^{\mu} S_{i}^{\alpha} m_{\alpha}^{\mu} \right]\right\} 
\label{e18}.
\end{eqnarray}
where $Dm_{\alpha}^{\mu(\nu)}={\displaystyle \prod_{\mu(\nu)}}  {\displaystyle   
\prod_{\alpha}} dm_{\alpha}^{\mu(\nu)}/\sqrt{2\pi}$.

In this work, the structure of solutions for auxiliary fields
$m_{\alpha}^{\mu}$'s and $m_{\alpha}^{\nu}$'s is the same as in { Ref.} \cite{Amit2}.
We assume that the relevant contributions come from $m_{\alpha}^{\nu}$ { which are order unity while 
$m_{\alpha}^{\mu}$ is of order $1/\sqrt{N}$.} 
Therefore, the average over the $p-s$ independent random variables $\xi_{i}^{\mu}$
can be done using Eq. (\ref{e5}) which results { in:}
\begin{eqnarray}
\langle\langle \exp \left[ \beta J
\sum_{\mu=s}^{p}\sum_{\alpha}(\sum_{i}\xi_{i}^{\mu}S_{i}^{\alpha}) 
m_{\alpha}^{\mu}\right]\rangle\rangle_{\xi}=
\exp\left[\sum_{i}\sum_{\mu=s}^{p}\ln(
\cosh(\beta J\sum_{\alpha}S_{i}^{\alpha} m_{\alpha}^{\mu}))
\right]
\label{equa19}.
\end{eqnarray}

{ The 
argument of the exponential in the} 
right hand side of 
Eq. (\ref{equa19}) can
be expanded up to second order in $m_{\alpha}^{\mu}$ 
The result is a quadratic term
of the spins variables $S_{i}^{\alpha}$ in the 
last exponential of 
Eq.
(\ref{e18}). 
This term can be linearized by introducing the spin glass order parameter
$q_{\alpha\beta}$ using the integral representation of the delta
function as we have done with the Kondo order parameter, { so
\begin{equation}
1=\int^{\infty}_{-\infty}\prod_{\alpha\beta}\delta
(q_{\alpha\beta}- \frac{1}{N}\sum_{i}S^{\alpha}_{i}S^{\beta}_{i})
\label{newsgq}.
\end{equation}
where $q_{\alpha\beta}$ is equivalent to the usual 
spin glass order parameter introduced in the classical 
SK model \cite{SK} which gives the transition to 
spin glass phase when $m_{\alpha}^{\mu}=0$. }
 
Therefore, after rescaling  $m_{\alpha}^{\mu}\rightarrow m_{\alpha}^{\mu}/\sqrt{N}$, 
{ the exponential involving $m_{\alpha}^{\mu}$ in Eq. (\ref{e18}) can be written as:}
\begin{eqnarray}
\exp\left\{\beta J N
\sum_{\mu=s}^{p}\sum_{\alpha}\left[-\frac{1}{2}(m_{\alpha}^{\mu})^2
+\frac{1}{N}(\sum_{i}\xi_{i}^{\mu}S_{i}^{\alpha})m_{\alpha}^{\mu}\right]\right\}
\nonumber\\=
\frac{1}{2\pi}\int_{-\infty}^{\infty} (\prod_{\alpha\beta} dq_{\alpha\beta}
d\bar{r}_{\alpha\beta})
\exp\left\{ -\frac{\beta J}{2} \sum_{\mu=s}^{p} m_{\alpha}^{\mu}\Lambda_{\alpha\beta} 
m_{\beta}^{\mu} \right.\nonumber\\
\left. i\sum_{\alpha\beta}\bar{r}_{\alpha\beta}\left( 
q_{\alpha\beta} -\frac{1}{N}\sum_{i}S_{i}^{\alpha} S_{i}^{\beta}\right)\right\}
\label{e20},
\end{eqnarray}
\noindent
where the matrix element
\begin{eqnarray}
\Lambda_{\alpha\beta}=\delta_{\alpha\beta} - 
\beta J q_{\alpha\beta} 
\label{e21}~.
\end{eqnarray}
Introducing Eqs. (\ref{e20}) and (\ref{e21}) into Eq. (\ref{e18}), the
$m_{\alpha}^{\mu}$ fields can be integrated to give:
%
\begin{eqnarray} 
\langle\langle \exp A_{SG}^{stat} \rangle\rangle_{\xi}=
\exp\left[
-\frac{\beta J \ p}{2N}{\displaystyle \sum_{\alpha=1}^n}{\displaystyle \sum_{i=1}^N}(S_{i}^{\alpha})^{2} \right]
\langle\langle \int^{+\infty}_{-\infty}  Dm_{\alpha}^{\nu}              
\nonumber\\
\exp\left\{\beta J N\sum_{\nu=1}^{s-1}\sum_{\alpha} 
\left[ -\frac{1}{2}(m_{\alpha}^{\nu})^{2}+
\right.\right.
\left.\left.
\frac{1}{N}(\sum_{i}\xi_{i}^{\nu}S_{i}^{\alpha}) m^{\nu}_{\alpha}\right]\right\} 
\rangle\rangle_{\xi} 
\int^{+\infty}_{-\infty}
{\displaystyle (\prod_{\alpha\beta}}\frac{dq_{\alpha\beta}\overline{r}_{\alpha\beta}}{2\pi})
\nonumber\\
\exp\{i\sum_{\alpha\beta}\overline{r}_{\alpha\beta}(q_{\alpha\beta}-\frac{1}{N}{\displaystyle \sum_{i=1}^{N}}S^{i}_{\alpha}S^{i}_{\beta})-
\frac{1}{2}(p-s)Tr\ln\underline{\underline{\Lambda}}) \}.
\label{e22}
\end{eqnarray}
%
Therefore, the averaged partition function (see Eq. (\ref{e17})) is obtained from Equations
(\ref{e13}-\ref{e151}) and (\ref{e22}) as
\begin{eqnarray} 
\langle\langle Z \rangle\rangle_{\xi}=
\int^{+\infty}_{-\infty}(\prod_{\alpha\nu}dm_{\alpha}^{\nu}) 
\int^{+\infty}_{-\infty}{\displaystyle 
(\prod_{\alpha\beta}}\frac{dq_{\alpha\beta}\overline{r}_{\alpha\beta}}{2\pi})
\times\nonumber \\ 
\exp\{-\beta N[\frac{J}{2}\sum_{\alpha}(m_{\alpha}^{\nu})^{2}+\frac{p-s}{2N\beta} 
Tr\ln\underline{\underline{\Lambda}}-\nonumber \\ 
\frac{i}{N\beta}\sum_{\alpha}\overline{r}_{\alpha\alpha}q_{\alpha\alpha}- \frac{i}{N\beta}
\sum_{\alpha\neq\beta}
\overline{r}_{\alpha\beta}q_{\alpha\beta}]\}
\langle\langle\Omega\rangle\rangle_\xi
\label{e23}
\end{eqnarray}
where
\begin{eqnarray} 
\Omega=
\int D(\psi^{\ast}\psi)
\exp
\{ 
-i\sum_{\alpha\neq\beta}\overline{r}_{\alpha\beta}(\frac{1}{N}\sum_{i}S_{i}^{\alpha}S_{i}^{\beta})-
\nonumber\\
\sum_{\alpha}(\frac{\beta J p}{2}+i\overline{r}_{\alpha\alpha})
\frac{1}{N}\sum_{i}
(S_{i}^{\alpha})^{2} 
+
\beta J(\sum_{i}\xi_{i}^{\nu}S_{i}^{\alpha})
m^{\nu}_{\alpha}
+\nonumber\\ 
\sum_{ij}\sum_{\sigma\alpha}\sum_{\omega}
\psi_{i\sigma\alpha}^{\ast}(\omega)
g_{ij}^{-1}(\omega) \psi_{j\sigma\alpha}(\omega) 
\}
\label{e25}
\end{eqnarray}

The free energy (see Eq. \ref{e17}) is evaluated at the saddle point { and is} given by the condition that 
the first derivatives of integrating variables are zero. Therefore, for instance, we have
\begin{eqnarray} 
-i\overline{r}_{\alpha\alpha}=\frac{\beta^{2}J^{2}}{2}\langle \sum_{\nu}(m_{\alpha}^{\nu})^{2}\rangle=\frac{\beta^{2}J^{2}}{2}p\ r_{\alpha\alpha}  
\label{e27}
\end{eqnarray}
and
\begin{eqnarray} 
-i\overline{r}_{\alpha\beta}=\frac{\beta^{2}J^{2}}{2}\langle \sum_{\nu}(m_{\alpha}^{\nu}m_{\beta}^{\nu})\rangle=
\frac{\beta^{2}J^{2}}{2}p\ r_{\alpha\beta};~~ \alpha \neq \beta. 
\label{e27d}
\end{eqnarray}
The parameter $m_{\alpha}^{\nu}$ has been found as in the section I.
The problem is treated assuming the replica symmetric ansatz, therefore the order parameters are 
$q_{\alpha\beta}=q$, $q_{\alpha\alpha}=\overline{q}$, $r_{\alpha\alpha}= \overline{r}$, $r_{\alpha\beta}= r$ and 
$m_{\alpha}^{\nu}= m^{\nu}$. On the other hand, the trace of the matrix $\underline{\underline{\Lambda}}$ is obtained 
in terms of its eigenvalues \cite{Amit1} (see Eq. {\ref{e21}}). In consequence, the free energy can be written as  
\begin{eqnarray} 
\beta f=2\beta J_{K}\mid \lambda \mid^{2}+\frac{\beta J}{2}\sum_{\nu}(m^{\nu})^{2}
-\frac{a}{2}(\frac{\beta J q}{1-\beta J (\overline{q}-q)})\nonumber \\ 
+\frac{a}{2}\ln[1-\beta J(\overline{q}-q)] +\frac{\beta^{2} J^{2} a}{2}\overline{r}\ \overline{q}-
\frac{\beta^{2} J^{2} a}{2}r\ q-\nonumber \\
\lim_{n\rightarrow 0} \frac{1}{nN}\ln[
\langle\langle \Omega(r, \overline{r}, m^{\nu}, \mid \lambda\mid)\rangle\rangle_{\xi}]  ~~~~~~~~
\label{e28u}
\end{eqnarray}
{ where $a=p/N$.}

The sum over replica index produces quadratic forms into the function 
$\Omega(r, \overline{r}, m^{\nu}, \mid \lambda \mid )$. This term can be 
linearized by a Hubbard-Stratonovich 
transformation \cite{Alba1, Magal1, Magal2} 
where new auxiliary fields are introduced in the problem. Therefore, we have:
\begin{eqnarray} 
\Omega(r, \overline{r}, m^{\nu}, \mid \lambda \mid )=
\nonumber\\
\int^{+\infty}_{-\infty}  {\displaystyle \prod_{j=1}^{N}Dz_{j} }            
\int^{+\infty}_{-\infty}{\displaystyle \prod_{j=1}^{N}\prod_{\alpha=1}^{n}
Dw_{j}^\alpha }                    
\int D(\psi^{\ast}\psi)
\nonumber\\
\exp[\sum_{ij}\sum_{\alpha,\sigma}\sum_{\omega}\psi_{i\sigma\alpha}^{\ast}(\omega)
G_{ij}^{-1}(\omega)
\psi_{j\sigma\alpha}(\omega)]
\label{e28}
\end{eqnarray}
with $Dz=\frac{dz}{\sqrt{2\pi}}e^{-z^{2}/2}$ and  
\begin{eqnarray} 
G_{ij}^{-1}(\omega)=g_{ij}^{-1}(\omega)-\sigma(\overline{h}^{\alpha}_{i}(r, \overline{r})
+\beta\sum_{\nu}\xi_{i}^{\nu}m^{\nu})\delta_{ij}
\label{e29}
\end{eqnarray}
where $g_{ij}^{-1}(\omega)$ is given { by} 
Eq. (\ref{e14}) and { the} 
{ local spin glass component of the internal field}
\begin{eqnarray} 
\overline{h}^{\alpha}_{i}(r, \overline{r})=\sqrt{\beta J
a[\beta J( \overline{r}-r)-1]}w^{\alpha}_{i}+\beta J\sqrt{a r}z_{i}.
\label{e30}
\end{eqnarray}

The functional integral in Eq. (\ref{e28}) can be performed \cite{Magal1, Alba1, Magal2}, 
so we get
\begin{eqnarray} 
\Omega(r, \overline{r}, m^{\nu}, \mid \lambda \mid )=
\int^{+\infty}_{-\infty}                                             
(\displaystyle \prod_{j=1}^{N}Dz_{j})
\nonumber\\ 
\int^{+\infty}_{-\infty}                                               
(\displaystyle \prod_{j=1}^{N}\prod_{\alpha=1}^{n}Dw_{j}^{\alpha})
 \exp\left[ \sum_{\omega}\sum_{\alpha,\sigma}\ln[\det G_{ij}^{-1}(\omega)]\right] 
\label{e31}
\end{eqnarray}

The local field $\overline{h}^{\alpha}_{i}(r, \overline{r})$ applied in the $n$ replicated lattices is, 
in fact, associated with the replica diagonal and 
non-diagonal spin glass order parameters. 
The presence of the local field is
the fundamental technical issues which 
must be solved in order to proceed the calculations.
Therefore, we consider the following decoupling \cite{Magal1, Alba1, Magal2}:
\begin{eqnarray} 
\ln[\det G_{ij}^{-1}(\omega)]\approx \frac{1}{N}\sum_{i}
\ln[\det\Gamma_{\delta\nu\, \sigma}(\omega, \overline{h}^{\alpha}_{i}(r, \overline{r}), 
\xi_{i}^{\nu} )]
\label{e32}
\end{eqnarray}
which means that a constant field is applied in a fictitious Kondo lattice, so the problem can be solvable by 
a Fourier transform. 
The particular form of the decoupling is also useful because it allows the use of self-averaging property  
$\frac{1}{N}\sum_{i}f(\xi_{i})=\langle\langle f(\xi)\rangle\rangle_{\xi}$ \cite{Amit1},  which is 
valid in thermodynamic 
limit for finite $s-1$, the upper value of $\nu$. Actually, from now on, it is assumed that $s=2$. 
The self-averaging in 
Eq. (\ref{e32}) 
allows us to drop the 
{ site index $i$}
in Eq. (\ref{e31}).

The resulting sum over $k$ in Fourier transformed 
$\Gamma_{k, \sigma}(\omega, \overline{h}^{\alpha}(r, \overline{r}), \xi)$ can be replaced by an integral using a constant 
density of states for the conduction electrons $\rho(\epsilon)=\frac{1}{2 D}$ for $-D<\epsilon<D$. On the other hand, 
the sum over the Matsubara's frequencies in Eq. (\ref{e31}) can be performed following closely the procedure given in { Refs.} 
\cite{Magal1, Alba1, Magal2}. Finally, the free energy can be found as   
%
\begin{eqnarray} 
\beta f=2\beta J_{K}\mid \lambda \mid^{2}+
\frac{\beta J}{2}(m^{1})^{2}+\frac{a}{2}\ln[1-\beta J(\overline{q}-q)]-
\frac{a}{2}\frac{\beta J  q}{1-\beta J(\overline{q}-q)}+
\nonumber\\
\frac{\beta^{2} J^{2} a}{2}\overline{r}\ \overline{q}-\frac{\beta^{2} J^{2} a}{2}r\ q-
\int^{+\infty}_{-\infty} Dz\langle\langle \ln\{ \int^{+\infty}_{-\infty} Dw 
\exp[ \frac{1}{\beta D}\int^{+\beta D}_{-\beta D} dx 
\times \nonumber\\
 \ln[2\cosh(\frac{x+h(r,\overline{r},\xi)}{2})]+
 2\cosh(\sqrt{\Delta(r,\overline{r},\xi)^{2} +(\beta J_{K}
 \mid\lambda\mid)^{2}}  ]  \}\rangle\rangle_{\xi}
\label{e33}
\end{eqnarray}
%
where 
\begin{eqnarray} 
\Delta(r,\overline{r},\xi)=\frac{x-h(r,\overline{r},\xi)}{2}
\label{e34}
\end{eqnarray}
and
%
\begin{equation}
h(r,\overline{r},\xi)=
\sqrt{\beta J a[\beta J(\overline{r}-r)-1]}w + 
\beta J\sqrt{a r}z+\beta Jm^{1}\xi
\label{e35}
\end{equation}
%
is { a long range internal field} composed by two parts, a spin glass { one} already introduced in Eq. 
(\ref{e30}) 
and  { other one} 
associated with the order parameter $m^{1}$. 
This result can be compared with { Ref.} \cite{Magal1}.
In that work, a random Gaussian inter-site 
coupling among the localized spins has been used. The equivalent 
{ internal} field found there can be also decomposed in two parts: 
a ferromagnetic { one}, and a spin glass term associated with $q$ and the static 
susceptibility $\chi=\beta(\overline{q}-q)$. However, 
the dependence { of the internal field}  with $q$ and $\overline{q}$ is entirely 
distinct { in the present work} as it will be shown below.

The coupled set of equations for the order parameters can be found from Equation (\ref{e33})~
 using 
the saddle point conditions. In particular, there is a relationship between the following order parameters:
\begin{eqnarray}
r=\frac{q}{[1-\beta J(\overline{q}-q)]^{2}}~~,
\label{e36}
\end{eqnarray}
\begin{eqnarray}
r-\overline{r}= \frac{1}{\beta J}\frac{1}{[1-\beta J(\overline{q}-q)]}~~.
\label{e37}
\end{eqnarray}
{ If we replace Eqs. (\ref{e36}-\ref{e37}) in Eqs.(\ref{e33}-\ref{e35}), the theory becomes explicitly dependent on 
the $q$ and $\overline{q}$.}
Therefore, the minimum set of order parameters to be solved in order to obtain a global 
phase diagram is the spin glass order 
parameters $q$, $\overline{q}$ (related to diagonal matrix $q_{\alpha\alpha}$), the Kondo order parameter $\mid \lambda\mid$ 
and the $m^{1}=m$. 
The average over $\xi$ in Eq. (\ref{e33}) can be now performed using the parity properties of the  
functions dependent on 
$z$ and $w$. Therefore, the dependence on $\xi$ can be dropped. The remaining set of coupled equations for the  
order parameters 
are given by the corresponding saddle point equations.
\begin{figure}[t] 
\includegraphics[angle=270, width=8.5cm]{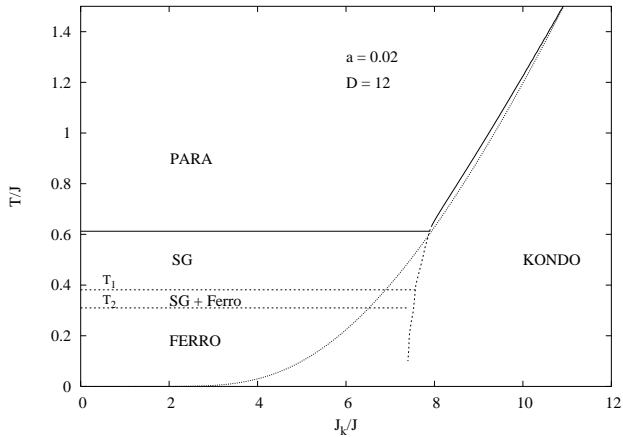}
\caption{ The phase diagram with $T/J$ {\it versus} $J_{K}/J$ for $a=0.02$ showing the phases SG (spin glass), FERRO (ferromagnetism) 
and KONDO (the Kondo state). In the region SG+Ferro, there is coexistence between SG 
and FERRO. The dotted line means 
the ``pure" Kondo temperature.}
\label{f1}
\end{figure}

\section{{ Discussions}}
In this paper, it has been investigated the competition between ferromagnetism and spin glass in a Kondo lattice model with   
a random { coupling between localized spins.
The coupling  is constructed (see Eq. (\ref{Jij}))
as a product of two random independent variables $\xi_{i}^{\mu}$ where $i=1...N$ (N number of sites) 
and $\mu=1...p$ which is a generalization of the Mattis model \cite{Mattis} for random magnetic systems. 
In the { strongly} frustrated limit 
$p\rightarrow\infty$ (when $N\rightarrow\infty$ and $p/N=1$), it is recovered a spin glass 
solution as S-K model 
\cite{Provost}.
The problem is solved 
using functional integral formalism, the static aproximation   and the  replica ansatz. The order parameters 
are obtained by combining methods 
proposed in { Refs.} \cite{Alba1}, \cite{Amit1} which  allow us to introduce an aditional 
parameter $a=p/N$ 
to control the level of frustration. }   
\begin{figure}[t] 
\includegraphics[angle=270, width=8.5cm]{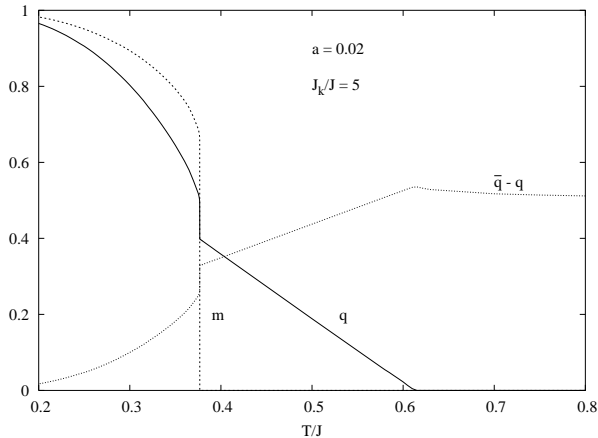}
\caption{The behavior of the order parameters $q$, $\overline{q}$ and $m$ showing the 
second order and the first order transition for $J_{K}=5J$.}
\label{f0}
\end{figure}

The numerical solutions of the order parameters $q$, $\overline{q}$, $m$ and 
$\mid \lambda\mid$ (from now on $\mid \lambda\mid=\lambda$ ) give as solutions the following thermodynamic phases: 
(a) a spin glass where $q\neq 0$ with $\lambda=0$ and $m=0$; 
(b) a Kondo state with $\lambda \neq 0$ and $q=0$ and $m=0$ ; 
(c) ferromagnetism which is given by the existence of the Mattis states described by 
$m \neq 0$ and $q\neq 0$ with $\lambda=0$. The solutions are displayed in diagrams $T/{J}$ {\it versus} 
$J_{K}/{J}$ for several values of $a$, where $T$ is the temperature, 
$J_{K}$ 
and ${J}$ are 
the strength of the intrasite Kondo coupling 
and the intersite random coupling between localized spins (see Eqs. (\ref{e2})-(\ref{e5})), respectively. 
Therefore, for a given $J_K/J$, it is possible to probe solutions 
for the order parameters equations $q$ 
, $\overline{q}$, $m$ 
and $\lambda$ 
in several situations ranging from weak 
frustration to strong 
one
just by varying the parameter $a$.  

{ The physical origin of the phases discussed in the previous paragraph can be understood from 
the model introduced in Eqs. (2)-(3) in which there are two interactions. The first one is 
on site Kondo coupling while the second one is the disordered coupling between localized spins. Therefore, it is
possible to identify several energy scales in the problem as $T$, $J_{K}$ and $J$. 
When temperature is high enough, there is only paramagnetism. As long as the Kondo energy scale becomes dominant in relation to the remaining ones, the emerging ordering is the complete screening of the localized spins in the 
whole lattice due to the Kondo effect. 
However, in a certain range of $J_K/J$, there are two possible magnetic orderings. In the spin glass one, the competing 
interactions between the spins can give rise to frustration in which there is a large number of degenerate states for the spins configurations. Therefore, 
there is no long range order correlation among spins orientations due to the presence of frustration which leads the spins to be frozen in random orientations.  
In contrast, in the ferromagnetic regime, the Mattis states correspond to the situation in which the spins stabilize aligned with 
the $\xi_{i}$'s due to the low frustration level. 
The prevalence of one or other regime discussed above depends on the parameter $a$ which controls the frustration level for the 
random coupling given in Eq. (1) as well as $T$.}

\begin{figure}[t] 
\includegraphics[angle=270, width=8.5cm]{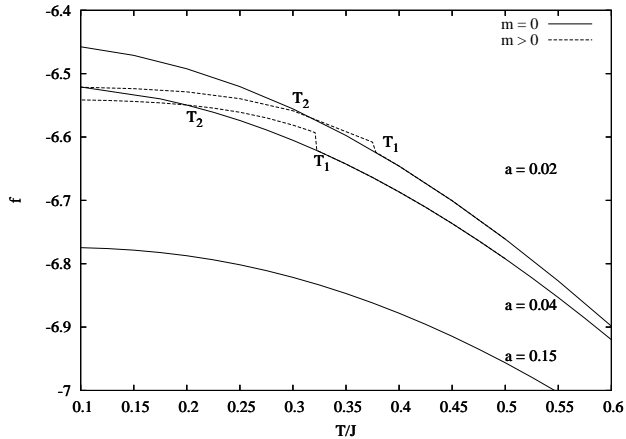}
\caption{The free energy as a function of temperature for $a=0.02$, $a=0.04$, and 
$a=0.15$ showing the region {(between $T_1$ and $T_2$)} where there are multiple solutions for the order parameters.}
\label{f01}
\end{figure}

It should be remarked that
previous investigation, using the same model, has  adopted the standard Gaussian random  
 coupling
\cite{Magal1, Alba1} (as the  
S-K model) for the intersite interaction between localized spins. Nevertheless,  in this  strongly frustrated 
approach { $T_{c}>T_{f}$ ($T_{c}$ and $T_{f}$ are the Curie and freezing temperatures, respectively)
in disagreement with  
experimental results 
for  $CeNi_{1-x}Cu_{x}$}. 
Therefore, the motivation for the present work is to understand better the role of disorder 
{ as source of frustration} in a Kondo lattice model and, { for that reason} 
to possibly address the experimental findings of $CeNi_{1-x}Cu_{x}$ phase diagram.

In Fig. 1, the results for $a=0.02$ are presented. 
For high 
$T/{J}$ 
and small $J_{K}/{J}$, the numerical solutions display a paramagnetic (PARA) behavior with $q=0$, $m=0$ and 
$\lambda=0$. The solutions remain the same in this 
small $J_{K}/{J}$ region { until} $T\approx 0.61{J}$, where $q$ starts to be 
continuously non-zero indicating a second order phase transition to a spin glass phase. The behavior of the 
order parameters as a function of the temperature is shown in Fig. 2.  
In particular, we can see that { at the same temperature} where $q\neq 0$, there is also 
a cusp in 
$\overline{\chi}=\chi/\beta=(\overline{q}-q)$ ($\chi$ is static susceptibility). 
If the temperature is lowered, the results remain yielding a spin 
glass solution until $T\approx 0.38{J}$. 
{  
From that point, the parameters 
$m$ and $q$ become simultaneously different from zero 
similar to the ferromagnetic solution already found  
in Ref. \cite{Magal1}. However, in the present case  there is an abrupt change in their behaviour indicating 
a
first order transition.   
In fact, for $0.31{J}\lesssim T \lesssim 0.38{J}$,
we have found 
metastable
solutions with $m\neq 0$ and $q\neq 0$ which corresponds to 
the Mattis states. 
The emergence of these metastable 
Mattis states at 
temperature 
$T_{1}$ 
can be seen clearly just by following the free energy (see Fig. 3). In Fig. 1, the corresponding 
region 
has been named as SG + Ferro. 
Finally, at 
$T_{2}$, 
the spin glass 
solutions become unstable thermodynamically while the solution 
with $m\neq 0$ and $q\neq 0$ (the Mattis states) becomes stable. }                          

{ When} $J_{K}/J$ is enhanced (see Fig. 1), 
it is found a line transition $J_{K_{c}}(T/J)$ for the Kondo state. 
This kind of solution had
already been found in { Refs.} \cite{Magal1, Alba1, Magal2}. 
The nature of this line transition is complex. It is second order 
at high temperature changing to first order at low temperature. However, there are evidences 
indicating that this complexity could be nonphysical, in fact, related to the approximations made in the present approach \cite{CoqAlba}.

If the parameter $a$ is increased (for instance $a=0.04$, in Fig. { 4}), we have, basically, a phase diagram displaying the same 
situation already shown in Fig. 1. However, the spin glass stability range is increased. 
{ Finally, for} $a=0.15$, {there is no more Mattis states as solution}, the spin glass { solution} is { entirely} 
dominating for 
{ $J_{K }<J_{Kc}(T)$ in which $J_{Kc}(T)$ is the phase boundary of the Kondo state.}
In this limit, it is recovered the results obtained in { Ref.} 
\cite{Alba1}.     

\begin{figure}[t] 
\includegraphics[angle=270, width=8.5cm]{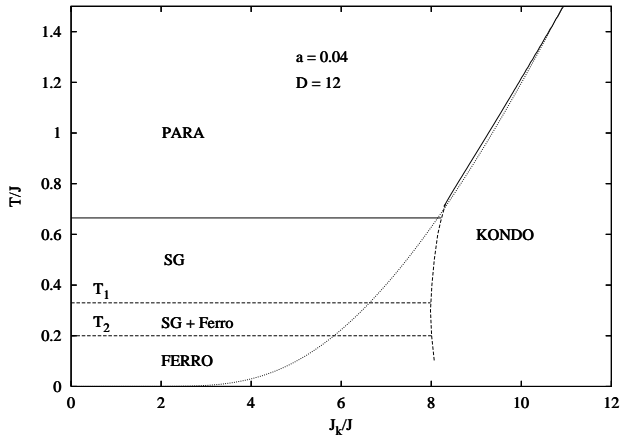}
\caption{ The phase diagram with $T/J$ {\it versus} $J_{K}/J$ for $a=0.04$ showing the phases SG (spin glass), FERRO (ferromagnetism) 
and KONDO (the Kondo state). In the region SG+Ferro, there is the coexistence between SG and FERRO. The dotted line means 
the ``pure" Kondo temperature.}
\label{f2}
\end{figure}
\begin{figure}[t] 
\includegraphics[angle=270, width=8.5cm]{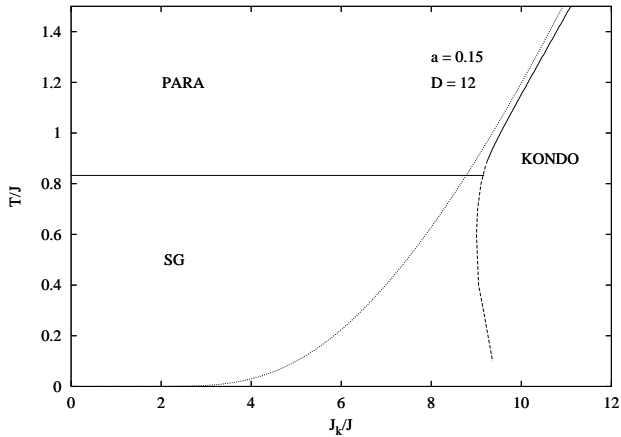}
\caption{ The phase diagram with $T/J$ {\it versus} $J_{K}/J$ showing the phase SG (spin glass). 
The dotted line means the ``pure" Kondo temperature.}
\label{f3}
\end{figure}

The experimental results for $CeNi_{1-x}Cu_{x}$  can be  now addressed.
The phase diagrams obtained with 
bulk 
methods \cite{Gomez-Sal1,Gomez-Sal2} and $\mu$SR technique \cite{Gomez-Sal3} 
show { that the chemical substitution of $Ni$ produces
in the alloy  
a complex interplay between 
 Kondo effect
and magnetism
where, for example, 
 the spin glass phase is always found 
at higher temperature than the 
ferromagnetic one 
(whatever the nature of the spin glass region is). Moreover, it also shows the increase of 
the freezing temperature  
with the decrease of $Cu$ content
in the 
alloy until 
$x \approx 0.4$. 
On the other hand, for $x\lesssim 0.2$ there is
a considerable reduction of $Ce$ localized magnetic moments  
due to the Kondo 
 effect.

If it is assumed that the substitution of $Cu$ by $Ni$ can be related to the parameters $J_{K}$ 
and $a$, some important aspects of the experimental scenario 
can be  reproduced, 
for example:  

i)  For $J_{K }<J_{Kc}(T)$ and small $a$ (weak frustration). 
In this regime the magnetic solutions are dominant  with no trace of a Kondo solution. However, the combination of high
temperature and the local complex randomness prevents any kind of stable alignement of 
the spins (represented by the order parameter $m$). 
On the contrary, the solutions for the order parameter show 
a continous direct transitions 
 from
paramagnetism 
 to
spin glass. Eventually, 
at lower temperature, the randomness is not enough to keep avoiding any stable alignement of the spins. 
Therefore, the Mattis states start to appear 
first as metastable solutions until becoming the thermodynamical stable solutions of the 
problem.  As consequence, in this weak frustration limit, the correct order for the transition temperatures 
is recovered as compared with the experimental situation when $0.8 \gtrsim x \gtrsim 0.4$, where 
$x$ is the content of $Cu$ \cite{Gomez-Sal2}.

As long as the level of frustration $a$ is increasing, the spin glass component 
of the internal field becomes dominant in a larger range of temperature. Therefore, 
it  is 
obtained 
an enlargement of the spin glass region (see Figure 1, 4 and 5) 
which precedes 
the onset of the ferromagnetism. 
As a consequence,  the freezing 
temperature $T_{f}$ 
is also increased,   which resembles  the experimental situation 
when $x\rightarrow 0.4$ \cite{Gomez-Sal2}. 

ii) When $J_{K} >J_{Kc}(T)$, 
 the magnetic solutions disappear and the Kondo state appears as the  unique 
solution in the problem  for any value of $a$ which means that there is a complete 
screening of localized spins due to the Kondo effect  as in the experimental results 
for the rich $Ni$ region
\cite{Gomez-Sal2}.  

 However, there is  still some disagreement related to the results obtained from $\mu$Sr 
spectroscopy \cite{Gomez-Sal3}  and specific heat \cite{Espeso}  which 
suggests the presence of nanoclusters which are frozen at some temperature \cite{Espeso}. 
For lower temperatures,  those 
clusters would percolate yielding 
a ferromagnetism \cite{Espeso} in a similar process 
proposed to explain manganites \cite{Dagotto}.    
Our results,  instead, in the  weak frustrated limit, indicate 
a continuous transition to a spin glass  phase 
and, then, at lower temperature, the 
presence of a first order one 
with a coexistence region between spin glass and  ferromagnetism
(given by the Mattis states). 
Nevertheless, we believe that the present mean field theory is a clear 
improvement 
in the sense that it  provides an effective mechanism which, at least, gives the correct 
ordering of the magnetic transition temperatures.

To conclude, in this work, it has been studied the 
spin glass-ferromagnetism-Kondo phase transitions
 in a Kondo 
lattice model with a random coupling $J_{ij}$ between the localized spins.  
The $J_{ij}$ is given as a product of random variables $\xi_{i}^{\mu}$ ($\mu=1...p$, $i=1...N$). 
This choice for $J_{ij}$ 
introduces a parameter $a=p/N$ which allows us to control the degree 
 of frustration. Thereby, 
 the balance between the two parameters $J_{K}/J$ 
and $a$ controls the emergence 
of the 
 different
solutions in the problem. 
For small $a$ (weak frustration) and $J_{K}/J$, 
we have found  that there is only
the presence  of
spin glass and ferromagnetic solutions,    
and, particularly that the freezing temperature is higher than  
the transition temperature 
where  
ferromagnetic solutions are found in good agreement with experiment in Ce(Ni,Cu) alloys. For large $J_K/J$, there is only a Kondo state, whatever the value of $a$ is.
{ The results obtained here are interesting for the study of the role of disorder, which seems to be better described by taking into account 
an average of discrete values $\xi$'s, rather than a direct average of the intersite $J$ values, as suggested also by new experimental results 
in $Ce$$Ni_{1-x}$$Cu_{x}$ \cite{Espeso} alloys. } 

{\bf Acknowledgments}

The numerical calculations were, in part,  
per\-for\-med at LSC (Cur\-so de Ci\-\^en\-cia da Com\-pu\-ta\-\c{c}\~ao, UFSM) and 
grupo de F\'\i\-sica Estat\'\i\-tica-IFM, Universidade Federal de Pelotas. The authors
are grateful to 
Prof. J. I. 
Espeso, Prof. J. C. 
Gomez-Sal and Prof. Alba Theumann. 
This work was partially supported by the Brazilian agencies FA\-PERGS 
(Fun\-da\-\c{c}\~ao\ de Am\-pa\-ro \`a Pes\-qui\-sa do Rio Gran\-de do Sul) 
and CNPq (Con\-se\-lho Na\-cio\-nal de De\-sen\-vol\-vi\-men\-to 
Ci\-en\-t\'\i\-fi\-co e Tec\-no\-l\'o\-gi\-co).

\end{document}